\documentclass[prb,twocolumn,showpacs,floatfix]{revtex4}
\usepackage{graphicx}
\usepackage{times}
\usepackage{bm}

\def\be{\begin{equation}}
\def\ee{\end{equation}}
\def\bea{\begin{eqnarray}}
\def\eea{\end{eqnarray}}
\def\bse{\begin{subequations}}
\def\ese{\end{subequations}}

\def\be{\begin{eqnarray}}
\def\ee{\end{eqnarray}}

\begin{document}

\title{Berry phase mediated topological thermoelectric transport in gapped
single and bilayer graphene}
\author{Chuanwei Zhang$^1$}
\author{Sumanta Tewari$^{2}$}
\author{S. Das Sarma$^{3}$}
\affiliation{$^{1}$Department of Physics and Astronomy, Washington State University,
Pullman, WA 99164\\
$^{2}$Department of Physics and Astronomy, Clemson University, Clemson, SC
29634\\
$^{3}$Condensed Matter Theory Center, Department of Physics, University of
Maryland, College Park, MD 20772}

\begin{abstract}
We consider the anomalous thermoelectric transport in gapped single and
bilayer graphene where the gap may be due to broken inversion symmetry. In
the presence of the gap, non-trivial Berry phase effects can be shown to
mediate a transverse thermoelectric voltage in response to an applied
temperature gradient even in the absence of a perpendicular magnetic field.
This spontaneous, anomalous Nernst effect is nonzero for non-uniform
chemical potential in the two inequivalent valleys in the graphene band
structure. Conversely, the Nernst response can be used to create a
valley-index polarization between the two transverse sample edges as in the
analogous valley Hall effect.
\end{abstract}

\maketitle

\section{Introduction}

Interesting electric transport properties of the Dirac quasiparticles of
graphene have recently come under focus.\cite%
{Geim,Neto,Novoselov,Novoselov2,Zhang} Graphene, which consists of a single
layer of carbon atoms in a bi-partite honeycomb lattice, has a unique band
structure that includes two inequivalent valleys in the first Brillouin zone
of the momentum space. In gapless graphene, the valleys are characterized by
the conduction and the valence bands touching each other at zero energy.
Near the valleys, the appropriate quasiparticle dispersion relations are
linear in momentum, similar to those of massless Dirac quasiparticles.
Inter-valley-scattering is generally suppressed \cite{Morozov,
Morpurgo,Gorbachev}, and will be ignored here, due to the absence of a large
enough scattering wavevector mixing the two Dirac points.

In some graphene samples, e.g., epitaxial graphene on SiC substrate, a
sizable gap has been reported \cite{Zhou} in the quasiparticle spectrum near
the Dirac points. Since the gap approaches zero with increasing sample
thickness, it has been attributed to the effects due to the substrate, which
is believed to break the sublattice (hence, the space inversion (SI))
symmetry.\cite{Zhou} Note that a quasiparticle gap in graphene can also
result from confined geometries such as in the graphene nanoribbons.\cite%
{Kim1,Sarma} In bilayer graphene, an applied interlayer voltage bias can
lead to a gap due to broken SI symmetry.\cite{Ohta} In such systems, the
effects of the momentum space Berry curvature,\cite{Berry} which is sharply
peaked at the two valleys, give rise to non-trivial, topological, electric
transport phenomena.\cite{Sinitsyn,Xiao1} In this paper, we discuss the
effects of the topological Berry phases on another class of transport
coefficients, the thermoelectric response coefficients. We will focus, in
particular, on the topological Nernst effect
of gapped graphene. In light of the recent successful measurements of the
regular thermoelectric coefficients in graphene,\cite{Wei,Kim,Ong} the
anomalous, topological Nernst response we discuss here acquires particular
relevance and importance. Although the specific calculations in this paper
apply to graphene samples with broken SI symmetry, they should, more
generally, be qualitatively valid in the presence of a gap of any origin
near the Dirac points. Note that an energy gap is ultimately essential for
use of graphene as an electronic material.

We calculate the Berry phase supported, spontaneous (\emph{i.e.,} it exists
even without an external magnetic field) Nernst effect of the low energy
quasiparticles in graphene, and show that it is non-zero for a non-zero
valley polarization in the sample. While a non-zero valley polarization can
be artificially created,\cite{Rycerz} it can also be induced by an applied
magnetic field itself in a generic Nernst measurement set up.\cite{Xiao1}
The regular Nernst effect in gapped graphene may, therefore, have a
substantial anomalous contribution because of the induced valley
polarization. Even in the absence of any valley polarization, the
spontaneous Nernst signals are non-zero from the individual valleys, but are
equal and opposite to each other so the total signal is zero. Therefore,
they can be used to create a non-zero valley polarization on the transverse
edges, which can be technologically important.

In recent experiments, thermoelectric coefficients, in particular, the
Nernst coefficient, have been measured for non-zero transverse magnetic
fields in gapless graphene.\cite{Wei,Kim,Ong} The topological Nernst effect
we discuss here exists even in the absence of any external magnetic field;
it is solely driven by the effective magnetic field, given by the Berry
curvature, near the valley centers in the momentum space. In the
thermoelectric experiments, a temperature gradient $-\mathbf{\nabla }T$,
applied along, say, the $\hat{x}$ direction produces a measurable transverse
electric field. The total charge current in the presence of the electric
field $\mathbf{E}$ and $-\mathbf{\nabla }T$ is given by $J_{i}=\sigma
_{ij}E_{j}+\alpha _{ij}(-\partial _{j}T)$, where $\sigma _{ij}$ and $\alpha
_{ij}$ are the electric and the thermoelectric conductivity tensors,
respectively. Setting $\mathbf{J}$ to zero, the Nernst signal, defined as $%
e_{N}\equiv E_{y}/\left\vert \mathbf{\nabla }T\right\vert =\rho \alpha
_{xy}-S_{xx}\tan \theta _{H}$, is measured, where $\alpha _{xy}$ is the
Nernst conductivity defined via the relation $J_{x}=\alpha _{xy}\left(
-\partial _{y}T\right) $ in the absence of the electric field, $\rho
=1/\sigma _{xx}$ is the longitudinal resistance, $S_{xx}=E_{x}/\left\vert
\mathbf{\nabla }T\right\vert =\rho \alpha _{xx}$ is the thermopower, and $%
\tan \theta _{H}=\sigma _{xy}/\sigma _{xx}$ is the Hall angle. If the second
term is small, which is sometimes the case,\cite{Ong2} $\rho \alpha _{xy}$
completely defines the Nernst signal, but in the most general case one
should extract $\alpha _{xy}$ from the experimental data to compare with our
calculations.

In this paper, the Berry phase mediated topological contribution to $\alpha
_{xy}$ is calculated through two different approaches. In the first, path
integral, method, we first compute the frequency-dependent off-diagonal
component of the electric conductivity tensor, $\sigma _{ij}(\Omega )$, for
the Dirac quasiparticles in graphene. At low temperatures, the Nernst
conductivity $\alpha _{xy}$ can be related to the zero temperature DC Hall
conductivity $\sigma _{xy}$ through the Mott relation,\cite{Marder} Eq.~(\ref%
{Mott}). This method has the advantage that it produces, as a by-product,
the imaginary part of the AC Hall conductivity, $\sigma _{xy}^{\prime \prime
}$, which can be related to the possible polar Kerr effect \cite%
{Victor,Tewari} in graphene. Indeed, in the case of a non-zero valley-index
polarization in the sample, the Berry phase induced polar Kerr effect will
be non-zero and observable. Another advantage of this method is that it can
give a neat analytical expression of the Nernst conductivity at low
temperatures. The effects of quasiparticle nonlinear dispersion relations on
the topological transport properties of graphene have also been discussed in
this framework. This method, however, is approximate, and gives the values
of the Nernst coefficient only at low enough temperatures. In a more general
method, which is valid also at higher temperatures, the anomalous
contribution to $\alpha _{xy}$ can be calculated from the coefficient $\bar{%
\alpha}_{xy}$, which determines the transverse heat current $\mathbf{J}^{h}$
in response to an electric field $\mathbf{E}$: $J_{x}^{h}=\bar{\alpha}%
_{xy}E_{y}$. The coefficients are related by one of the Onsager relations: $%
\bar{\alpha}_{xy}=T\alpha _{xy}$.\cite{Xiao2,Cooper} This way of computing $%
\alpha _{xy}$ makes use of the Berry curvature in the graphene band
structure (see Eq. (\ref{entropy})) more directly. It is a much more general
method, which we have used in our plots for the Nernst coefficient as a
function of temperature in Figs. 2 and 3. Using this method, we have been
able to go beyond the linear temperature dependence arising out of the Mott
relation.

The paper is organized as follows: Sec. II gives the Dirac-like Hamiltonian
for single layer graphene. Section III is devoted to the topological Hall
conductivity of single layer graphene. We calculate the AC Hall conductivity
$\sigma _{ij}(\Omega )$ using the path integral approach and discuss the
possible polar Kerr effect for single layer graphene. In the limit $\Omega
=0 $, we obtain the DC Hall conductivity. The effects of nonlinear
quasiparticle dispersion relation on the Hall conductivity are also
discussed. In Sec. V, we calculate the topological Nernst conductivity $%
\alpha _{xy}$ via both the Mott relation and the transverse heat current
coefficient. We do this for both single and bilayer graphene. Finally,
Section VI is devoted to the conclusion.

\section{Dirac-like Hamiltonian}

In the tight-binding approximation, graphene single layer with staggered
sublattice potential (which breaks the space inversion symmetry) can be
modeled with the nearest-neighbor hopping energy $t$ and a site energy
difference $\Delta $ between the sublattices ,\cite{Nilsson}
\begin{equation}
H=\left(
\begin{array}{cc}
\Delta /2 & \zeta \left( \mathbf{p}\right) \\
\zeta ^{\ast }\left( \mathbf{p}\right) & -\Delta /2%
\end{array}%
\right) ,  \label{Hamiltonian1}
\end{equation}%
where
\begin{eqnarray}
\zeta \left( \mathbf{p}\right) &=&-t\sum_{i}e^{i\mathbf{p}\cdot \mathbf{%
\delta }_{i}}  \nonumber \\
&=&-te^{ip_{x}a/2}\left[ 2\cos \left( \frac{\sqrt{3}p_{y}a}{2}\right)
+e^{-i3p_{x}a/2}\right]  \label{parameter}
\end{eqnarray}%
with $\mathbf{\delta }_{1,2}=\frac{a}{2}\left( 1,\pm \sqrt{3}\right) $, $%
\mathbf{\delta }_{3}=a\left( -1,0\right) \hat{x}$, $a$ is the lattice
constant. The Hamiltonian (\ref{Hamiltonian1}) operates on the two-component
spinor $\hat{\Psi}_{\mathbf{p}}=(\hat{c}_{A\mathbf{p}},\hat{c}_{B\mathbf{p}%
}) $, where $\hat{c}_{A\mathbf{p}},\hat{c}_{B\mathbf{p}}$ denote the
annihilation operators on the two sublattices A, B. The most interesting
physics occurs around two valleys located at the Brillouin zone corners $%
\mathbf{K}_{1,2}=\frac{4\pi }{3\sqrt{3}a}\left( 0,\tau _{z}\right) $, where $%
\tau _{z}=\pm 1$ labels the two valleys. Denoting $\mathbf{k}=\mathbf{p-K}%
_{1,2}$ to measure the momentum from the valley centers, we can rewrite the
Hamiltonian (\ref{Hamiltonian1}) as
\begin{equation}
H=\frac{\Delta }{2}\sigma _{z}+\zeta _{1}\left( \mathbf{k}\right) \sigma
_{x}+\zeta _{2}\left( \mathbf{k}\right) \sigma _{y}.  \label{Hamiltonian2}
\end{equation}%
Here
\begin{eqnarray}
\zeta _{1}\left( \mathbf{k}\right) &=&-t\left[
\begin{array}{c}
2\cos \frac{k_{x}a}{2}\cos \left( \frac{2\pi }{3}\tau _{z}+\frac{\sqrt{3}%
k_{y}a}{2}\right) \\
+\cos \left( k_{x}a\right)%
\end{array}%
\right] ,  \label{term1} \\
\zeta _{2}\left( \mathbf{k}\right) &=&t\left[
\begin{array}{c}
2\sin \frac{k_{x}a}{2}\cos \left( \frac{2\pi }{3}\tau _{z}+\frac{\sqrt{3}%
k_{y}a}{2}\right) \\
-\sin \left( k_{x}a\right)%
\end{array}%
\right] .  \label{term2}
\end{eqnarray}%
The spectrum of the Hamiltonian consists of two branches with the
eigenenergies $E_{\pm }(\mathbf{k})=\pm \Lambda (\mathbf{k})$, where
\begin{eqnarray}
\Lambda (\mathbf{k}) &=&\left[ t^{2}\left( 1+4\cos ^{2}\Theta +4\cos \Theta
\cos \frac{3k_{x}a}{2}\right) +\frac{\Delta ^{2}}{4}\right] ^{\frac{1}{2}},
\label{Energy1} \\
\Theta &=&\frac{2\pi }{3}+\frac{k_{y}a\sqrt{3}}{2}\tau _{z}.  \nonumber
\end{eqnarray}%
Close to the valley centers, we can expand $\zeta _{1}\left( \mathbf{k}%
\right) $ and $\zeta _{2}\left( \mathbf{k}\right) $ to the first order in $%
\mathbf{k}$, and rewrite the Hamiltonian as%
\begin{equation}
\hat{H}=\frac{\Delta }{2}\sigma _{z}+\frac{3at}{2}\left( \tau
_{z}k_{y}\sigma _{x}-k_{x}\sigma _{y}\right) =\mathbf{\Lambda }\left(
\mathbf{k}\right) \cdot \mathbf{\sigma },  \label{Hamiltonian}
\end{equation}%
with $\Lambda _{1}=3at\tau _{z}k_{y}/2$, $\Lambda _{2}=-3atk_{x}/2$, $%
\Lambda _{3}=\Delta /2$. The energy spectrum becomes
\begin{equation}
\Lambda (\mathbf{k})=|\mathbf{\Lambda }(\mathbf{k})|=\frac{\sqrt{\Delta
^{2}+9a^{2}t^{2}k^{2}}}{2}.  \label{Energy}
\end{equation}

\section{Hall conductivity}

\subsection{AC Hall conductivity}

Introducing the Lagrangian density, $\mathcal{L}=i\omega _{n}\hat{I}-\hat{H}(%
\mathbf{k})$, where $\omega _{n}$ is a fermionic Matsubara frequency, we
obtain the electron Green's function (for a single valley) as a function of $%
\vec{k}=(i\omega _{n},k_{x},k_{y})$
\begin{equation}
G_{0}(\vec{k})=\mathcal{L}^{-1}=\frac{i\omega _{n}\,\hat{I}+\mathbf{\Lambda }%
(\mathbf{k})\cdot \hat{\mathbf{\sigma }}}{g(\omega _{n},\mathbf{k})}.
\label{Green-Function}
\end{equation}%
Here $g(\vec{k})=\left( i\omega _{n}\right) ^{2}-|\mathbf{\Lambda }(\mathbf{k%
})|^{2}$. In order to calculate the AC Hall conductivity, we consider the
electromagnetic potential $\vec{A}=(A_{0},A_{x},A_{y})$, where $A_{0}$ is
the scalar potential, and $\mathbf{A}=(A_{x},A_{y})$ is the vector
potential. Expanding $\mathcal{L}$ to the first order in $\vec{A}$, and
integrating out the fermions, we identify the coefficient of the
Chern-Simons-like term in the effective action as the anomalous AC Hall
conductivity for a single valley,\cite{Tewari}%
\begin{eqnarray}
&&\sigma _{xy}(\Omega _{m})=\sum_{\vec{k}}\frac{\Upsilon (\mathbf{k})}{%
g(\omega _{n},\mathbf{k})\,g(\omega _{n}+\Omega _{m},\mathbf{k})},
\label{Hall-Conductivity} \\
&&\Upsilon (\mathbf{k})=-2\bm\Lambda \cdot \left[ \frac{\partial \bm\Lambda
}{\partial k_{x}}\times \frac{\partial \bm\Lambda }{\partial k_{y}}\right] .
\label{B-w}
\end{eqnarray}%
Here the electromagnetic potential $\vec{A}\left( {\vec{q}}\right) $ was
assumed to be a function of the Fourier variables $\vec{q}=(i\Omega
_{m},q_{x},q_{y})$, with $\Omega _{m}$ a bosonic Matsubara frequency, and we
took the limit $\bm q\rightarrow 0$ to arrive at Eq.~(\ref{Hall-Conductivity}%
). Substituting the expression for $\bm\Lambda (\mathbf{k})$ from the
Hamiltonian (\ref{Hamiltonian}) into the formula (\ref{B-w}), we find%
\begin{equation}
\Upsilon (\mathbf{k})=\frac{-9\tau _{z}a^{2}t^{2}\Delta }{4}.  \label{B}
\end{equation}

Summing over the fermionic Matsubara frequency $\omega _{n}$ in Eq.~(\ref%
{Hall-Conductivity}) and analytically continuing to the real bosonic
frequency, $i\Omega _{m}\rightarrow \Omega +i\delta $, where $\delta $ is a
positive infinitesimal, we find,
\begin{equation}
\sigma _{xy}(\Omega )=\int \frac{d^{2}\mathbf{k}}{(2\pi )^{2}}\frac{\Upsilon
(\mathbf{k})\;\{f\left[ E_{+}(\mathbf{k})-\mu \right] -f\left[ E_{-}(\mathbf{%
k})-\mu \right] \}}{\Lambda (\mathbf{k})\,[\Omega +i\delta -2\Lambda (%
\mathbf{k})]\,[\Omega +i\delta +2\Lambda (\mathbf{k})]}.  \label{s_xy}
\end{equation}%
Here $f(E-\mu )$ is the Fermi occupation function at temperature $T$ and $%
\mu $ is the chemical potential. The minus sign between the two Fermi
functions is due to the opposite sign of the Berry curvature for the
conduction and the valence bands of electrons. Since $\Lambda (\mathbf{k})=%
\sqrt{\Delta ^{2}+9a^{2}t^{2}k^{2}}/2$ and $\Upsilon (\mathbf{k})$ are
constants, we can define $\varepsilon =k^{2}/2$ and change the integration
to
\begin{equation}
\sigma _{xy}(\Omega )=\int_{0}^{\infty }\frac{d\varepsilon }{2\pi }\frac{%
\Upsilon F\left( \mu \right) }{E\,\left( \epsilon \right) [\Omega +i\delta
-2E\,\left( \epsilon \right) ]\,[\Omega +i\delta +2E\,\left( \epsilon
\right) ]},  \label{AC}
\end{equation}%
where $E\,\left( \epsilon \right) =\frac{1}{2}\sqrt{\Delta
^{2}+18a^{2}t^{2}\epsilon }$, $F\left( \mu \right) =\{f\left[ E\,\left(
\epsilon \right) -\mu \right] -f\left[ -E\,\left( \epsilon \right) -\mu %
\right] \}$.

It is clear that if the two valleys have the same chemical potential, $\mu
_{K_{1}}=\mu _{K_{2}}$, the total AC Hall conductivity $\sigma
_{xy}^{t}(\Omega )=\sigma _{xy}^{K_{1}}(\Omega )+\sigma _{xy}^{K_{2}}(\Omega
)=0$, because of the opposite signs of the function $\Upsilon $ for the two
valleys. However, the Hall current for an individual valley is nonzero.

The imaginary part of the AC Hall conductivity, $\sigma _{xy}^{\prime \prime
}(\Omega )$, comes from the pole in Eq.\ (\ref{s_xy}) at $\Omega =2E\,\left(
\epsilon \right) $. This pole represents vertical interband transitions
induced by the electromagnetic wave. Taking the imaginary part of (\ref{s_xy}%
), we find
\begin{equation}
\sigma _{xy}^{\prime \prime }(\Omega )=-\frac{\Upsilon e^{2}}{\pi \Omega
^{2}\hbar }\int_{0}^{\infty }d\epsilon \delta \lbrack \Omega -2E\,\left(
\epsilon \right) ]F\left( \mu \right) .  \label{s''}
\end{equation}%
Here we have already taken into account the spin degeneracy. At $T=0$, $%
F\left( \mu \right) =0$ if $E\,\left( \epsilon \right) <\mu $ and $F\left(
\mu \right) =1$ for $E\,\left( \epsilon \right) >\mu $. Therefore, if $%
\Omega <2\mu $ or $\Omega <\Delta $, $\sigma _{xy}^{\prime \prime }(\Omega
)=0$. If $\Omega \geq 2\mu $ and $\Omega \geq \Delta $, we have
\begin{equation}
\sigma _{xy}^{\prime \prime }(\Omega )=\frac{e^{2}\Delta }{2h\Omega }.
\label{ACHall}
\end{equation}%
At finite temperatures, we have $\sigma _{xy}^{\prime \prime }(\Omega )=%
\frac{\Delta e^{2}}{2h\Omega }F\left( \mu \right) $ if $\Omega \geq \Delta $%
. For unequal population of the quasiparticles in the two valleys, the
imaginary part of the AC Hall conductivity is non-zero. Since the polar Kerr
angle, $\theta _{K}$, which is the angle of rotation of the plane of
polarization of a linearly polarized light on normal reflection from the
sample, is directly proportional to $\sigma _{xy}^{\prime \prime }(\Omega )$
\cite{Victor,Tewari}, in this case, there will be a non-zero polar Kerr
effect in graphene.

\subsection{DC Hall conductivity}

Taking $\Omega =0$ in Eq. (\ref{s_xy}), we have for the DC Hall
conductivity,
\begin{equation}
\sigma _{xy}(0)=-\int \frac{d^{2}\mathbf{k}}{(2\pi )^{2}}\frac{\Upsilon
F\left( \mu \right) }{4\Lambda ^{3}(\mathbf{k})}.  \label{DCHall2}
\end{equation}%
Notice that
\begin{equation}
\Pi =-\frac{\Upsilon }{4\Lambda ^{3}(\mathbf{k})\,}=\frac{9a^{2}t^{2}\Delta
\tau _{z}}{2\left( \Delta ^{2}+9a^{2}t^{2}k^{2}\right) ^{3/2}}
\label{Berrycur}
\end{equation}%
is just the Berry curvature for an individual valley \cite{Xiao1}. We see
that the Berry curvatures for the two valleys differ only by a sign. At zero
temperature and $\mu =0$, $\sigma _{xy}(0)=-\tau _{z}/4\pi $. We should
multiply the result by two to take account of spin degeneracy. Restoring $e$
and $\hbar $, we have (for a single valley)
\begin{equation}
\sigma _{xy}(0)=-e^{2}\tau _{z}/h.  \label{DChall}
\end{equation}%
For nonzero $\mu >\Delta /2$, i.e., $\mu $ lying in the conductance band,
\begin{equation}
\sigma _{xy}(0)=-\tau _{z}e^{2}\Delta /2h\mu .  \label{DChall3}
\end{equation}%
As $\mu $ becomes very big, $\sigma _{xy}(0)$ again becomes zero.

\subsection{Nonlinear spectrum}

\begin{figure}[t]
\includegraphics[scale=0.38]{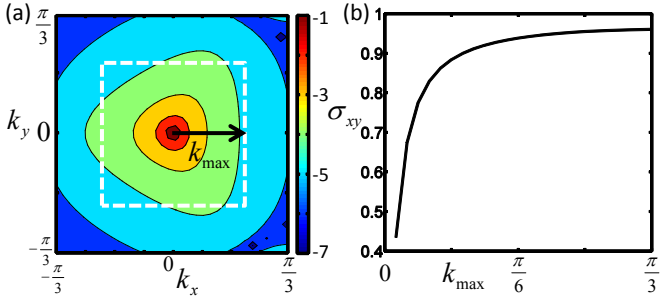}
\caption{(Color online) (a) Berry curvature $\Pi $ (we plot $-\Pi $ for
convenience) around the valley center $K_{2}$. $t=2.8eV$, $\Delta =0.26eV$.
The plot is in the logarithm scale. The dashed line indicates the
integration region for (b). (b) Plot of the DC conductivity with different
integration region $-k_{\max }\leq k_{x},k_{y}\leq k_{\max }$. The unit of $%
\protect\sigma _{xy}\left( 0\right) $ is $e^{2}/h$. }
\label{BC1}
\end{figure}

The form of the Berry curvature (\ref{Berrycur}) $\Pi =-\frac{\Upsilon }{%
4\Lambda ^{3}}$ also applies to the general Hamiltonian (\ref{Hamiltonian2})
without the linear approximation (i.e., the Hamiltonian (\ref{Hamiltonian}%
)). In this case, we should use $\Lambda $ determined by Eq. (\ref{Energy1}%
). It is easy to show that%
\begin{equation}
\Upsilon \left( \mathbf{k}\right) =-\frac{\sqrt{3}a^{2}t^{2}\Delta \tau _{z}%
}{2}\left[
\begin{array}{c}
2\cos (\frac{\pi }{6}+\frac{\sqrt{3}k_{y}a}{2}\tau _{z})\cos (\frac{3k_{x}a}{%
2}) \\
+\cos (\frac{\pi }{6}-\sqrt{3}k_{y}a\tau _{z})%
\end{array}%
\right] .  \label{Term4}
\end{equation}%
The lowest order in $\mathbf{k}$ (i.e., $\mathbf{k=0}$) of (\ref{Term4})
yields Eq. (\ref{B}). In the presence of the nonlinear part of the spectrum,
the symmetry relation between the Berry curvatures in the two valleys is
given by $\Pi _{K_{1}}\left( \mathbf{k}\right) =-\Pi _{K_{2}}\left( -\mathbf{%
k}\right) $, and the total Nernst conductivity will be non-zero only for
non-zero valley polarization. In Fig. \ref{BC1}a, we plot the Berry
curvature around the valley center $\mathbf{K}_{2}$ ($\tau _{z}=-1$). We see
that the Berry curvature is very small for a large $\mathbf{k}$. Therefore,
the large $\mathbf{k}$ regimes give very little contribution to the
transport quantity of graphene. In fact, for $\left( k_{x}a,k_{y}a\right)
=\left( \frac{\pi }{36},\frac{\pi }{36}\right) $, the Berry curvature is
only 1\% of its peak value. In Fig. \ref{BC1}b, we plot the DC Hall
conductivity at $T=0$ and $\mu =0$, taking account of the nonlinear
spectrum. Because of the nonlinear spectrum, we cannot take the integration
region up to $k=\infty $, the limit that is used for the linear spectrum.
However, the DC Hall conductivity $\sigma _{xy}(0)$ at each valley is still
very close to $-e^{2}\tau _{z}/h$ (the value in Eq. (\ref{DChall}) obtained
using the linear spectrum) for a reasonably large integration region.
Therefore the nonlinear spectrum does not play an important role on the
anomalous topological transport properties of graphene, and will be
neglected in the following.%

\section{Anomalous Nernst effect}

\subsection{Single layer graphene}

Since the transverse Hall conductivity can be non-zero for gapped graphene,
it is natural to ask whether there exists an anomalous Nernst effect. For
this purpose, at low temperatures, we can use the Mott relation \cite{Marder}%
,
\begin{equation}
\alpha _{xy}=-\frac{\pi ^{2}k_{B}^{2}}{3e}\frac{d\sigma _{xy}}{d\mu }T.
\label{Mott}
\end{equation}%
For $-\Delta /2<\mu <\Delta /2$, i.e., $\mu $ falling in the band gap, $%
\sigma _{xy}$ is independent of $\mu $, and we have $\frac{d\sigma _{xy}}{%
d\mu }=0$. Thus, $\alpha _{xy}=0$. If $\mu \geq \Delta /2$, we have
\begin{equation}
\frac{d\sigma _{xy}}{d\mu }=\frac{\tau _{z}e^{2}\Delta }{2h\mu ^{2}}.
\end{equation}%
Similarly, for $\mu \leq -\Delta /2$, $\frac{d\sigma _{xy}}{d\mu }=-\tau _{z}%
\frac{e^{2}}{h}\frac{\Delta }{2\mu ^{2}}$. We see that $\alpha _{xy}$ is
discontinuous at $\mu =\pm \Delta /2$.
\begin{figure}[t]
\includegraphics[scale=0.42]{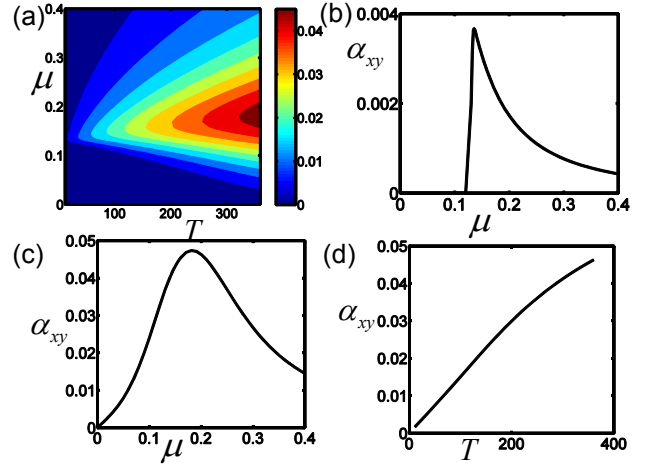}
\caption{(Color online) Anomalous Nernst signal for a single valley for
single layer graphene. The units for $\protect\mu $, $T$, and $\protect%
\alpha _{xy}$ are eV, K, and $k_{B}e/\hbar $, respectively. The parameters $%
t=2.82eV$, $\Delta =0.26eV$. (a) Contour plot for $\protect\alpha _{xy}$.
(b) Plot of $\protect\alpha _{xy}$ with respect to $\protect\mu $ for $%
T=11.6 $ K. Note the discontinuity at $\protect\mu =\Delta /2$, as predicted
by the Mott relation. The decay for $\protect\mu >\Delta /2$ follows $\sim 1/%
\protect\mu ^{2}$. (c) Plot of $\protect\alpha _{xy}$ with respect to $%
\protect\mu $ for $T=360$ K. Note the disappearance of the discontinuity.
(d) Plot of $\protect\alpha _{xy}$ with respect to $T$ for $\protect\mu =0.2$
eV.}
\label{single}
\end{figure}

\begin{figure}[t]
\includegraphics[scale=0.45]{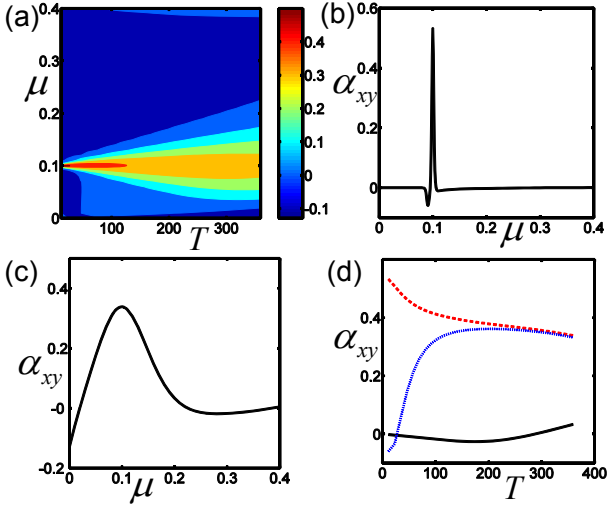}
\caption{(Color online) Anomalous Nernst signal for a single valley for
bilayer graphene. The units for $\protect\mu $, $T$, and $\protect\alpha %
_{xy}$ are eV, K, and $k_{B}e/\hbar $, respectively. The parameters $%
t=2.82eV $, $V=0.2eV$, $t_{\bot }=0.4eV$. (a) Contour plot for $\protect%
\alpha _{xy}$. (b) Plot of $\protect\alpha _{xy}$ with respect to $\protect%
\mu $ for $T=11.6$ K. Note the dramatic variation of $\protect\alpha _{xy}$
around $\protect\mu =V/2$. (c) Plot of $\protect\alpha _{xy}$ with respect
to $\protect\mu $ for $T=360$ K. (d) Plot of $\protect\alpha _{xy}$ with
respect to $T$ for $\protect\mu =0.2$ eV (Solid line), 0.1 eV (Dashed line),
0.091 eV (Dotted line). }
\label{bilayer}
\end{figure}

More generally, the anomalous contribution to $\alpha _{xy}$ can be
calculated from the transverse heat current coefficient $\bar{\alpha}_{xy}$.
This way of computing $\alpha _{xy}$ makes direct use of the Berry curvature
in the graphene band structure. In the presence of the Berry curvature and
the electric field, the electron velocity acquires the additional anomalous
term $\hbar \mathbf{v_{\mathbf{k}}}=e\mathbf{E}\times \Pi (\mathbf{k})$ \cite%
{Sundaram}. This velocity multiplied by the entropy density produces the
transverse heat current coefficient \cite{Zhang2}:
\begin{equation}
\bar{\alpha}_{xy}=T\alpha _{xy}=\frac{2e}{\beta \hbar }\sum_{n}\int \frac{%
d^{2}\mathbf{k}}{\left( 2\pi \right) ^{2}}\,\Pi _{n}(\mathbf{k})\,s_{n}(%
\mathbf{k}).  \label{entropy}
\end{equation}%
Here $s(\mathbf{k})=-f_{\mathbf{k}}\ln f_{\mathbf{k}}-(1-f_{\mathbf{k}})\ln
(1-f_{\mathbf{k}})$ is the entropy density of the electron gas, $f_{\mathbf{k%
}}=f[E_{n}(\mathbf{k})]$ is the Fermi distribution function, and the sum is
taken over both bands. This formula for $\alpha _{xy}$, which we use in the
figures \ref{single} and \ref{bilayer}, coincides with the expression
derived from Eq.~(\ref{Mott}) only at low temperatures. If $\mu _{K_{1}}=\mu
_{K_{2}}$, we see again that $\alpha _{xy}=0$, since the Berry curvature $%
\Pi _{K_{1}}=-\Pi _{K_{2}}$. However, the anomalous Nernst coefficient for
each valley remains non-zero. The anomalous Nernst effect, for a uniform
chemical potential, can therefore be used to create equal and opposite
valley-index polarization on the two transverse edges. This situation is
similar to the analogous spin Hall effect \cite{Nagaosa} or possible valley
Hall effect in graphene \cite{Xiao1}. For unequal chemical potential in the
valleys, which can be either artificially created \cite{Rycerz} or induced
by a perpendicular magnetic field in a generic Nernst set up, there will be
a non-zero anomalous contribution to the Nernst effect in graphene. The
unequal valley chemical potential in the presence of an external magnetic
field is due to the Zeeman-type coupling of the magnetic field with the
substantial magnetic moment associated with the valleys, which can be 30
times bigger than the usual spin magnetic moment \cite{Xiao1}.

Since Eq. (\ref{entropy}) is more general than the Mott relation, and may be
used to go beyond the linear temperature dependence of $\alpha_{xy}$ (\ref%
{Mott}), we numerically evaluate Eq. (\ref{entropy}) to calculate the Nernst
coefficient. We plot the anomalous Nernst coefficient as a function of $\mu $
and $T $ for a single valley in single layer graphene in Fig. (\ref{single}%
). In Fig. \ref{single}a, there is a nonzero Nernst coefficient even when
the chemical potential is inside the band gap, while the Mott relation
yields a zero Nernst signal in this case. In Fig. \ref{single}b, Nernst
conductivity $\alpha _{xy}$ increases sharply to a maximum, while the Mott
relation yields a step jump at $\mu =\frac{\Delta }{2}$. In Fig. \ref{single}%
c, the step jump is further smoothed out at medium temperatures (note that $%
T $ here is still much smaller than $\Delta $). The deviation of $\alpha
_{xy}$ from the linear temperature dependence given by the Mott relation is
clearly visible in Fig. \ref{single}d.

\subsection{Anomalous Nernst effect in bilayer graphene}

The Hamiltonian for a single valley in biased bilayer graphene can be
written as \cite{Nilsson}
\begin{equation}
\hat{H}=\left(
\begin{array}{cccc}
V/2 & tke^{i\phi _{\mathbf{k}}} & t_{\bot } & 0 \\
tke^{-i\phi _{\mathbf{k}}} & V/2 & 0 & 0 \\
t_{\bot } & 0 & -V/2 & tke^{-i\phi _{\mathbf{k}}} \\
0 & 0 & tke^{i\phi _{\mathbf{k}}} & -V/2%
\end{array}%
\right)
\end{equation}%
where $V$ is the interlayer bias voltage, $t_{\bot }$ and $t$ describe
interlayer and intralayer nearest neighbor hoppings and $\phi _{\mathbf{k}}$
is the angle of $\mathbf{k}$ on the 2D momentum space. The eigenenergies and
eigenstates can be obtained analytically, which yield the Berry curvature
for the \textit{n}-th band%
\begin{equation}
\Pi _{n}=\frac{1}{k}\frac{\partial }{\partial k}\left[ A_{n}^{2}\left(
\gamma _{1n}^{2}-\gamma _{3n}^{2}\right) \right] .  \label{Berry-bi}
\end{equation}%
Here $A_{n}=\left( 1+\gamma _{1n}^{2}+\gamma _{2n}^{2}+\gamma
_{3n}^{2}\right) ^{-1/2}$, $\gamma _{1n}=tk/\left( E_{n}-V/2\right) $, $%
\gamma _{2n}=-\left[ t^{2}k^{2}/\left( E_{n}-V/2\right) -\left(
E_{n}-V/2\right) \right] /t_{\bot }$, $\gamma _{3n}=\gamma _{2}tk/\left(
E_{n}+V/2\right) $, and, finally,
\begin{equation}
E_{n}=\pm \frac{1}{2}\sqrt{%
\begin{array}{c}
2t_{\bot }^{2}+V^{2}+4t^{2}k^{2} \\
\pm 2\sqrt{4\left( V^{2}+t_{\bot }^{2}\right) t^{2}k^{2}+t_{\bot }^{4}}%
\end{array}%
}
\end{equation}%
are the eigenenergies.

In Fig. \ref{bilayer}, we plot $\alpha _{xy}$ as a function of $\mu $ and $T$%
, similar as in Fig. \ref{single}. Note that, in this case, $\alpha _{xy}$
can change sign in some regimes of the parameter space. Such sign changes
are due to the complicity of the four bands of bilayer graphene. At low
temperatures, $\alpha _{xy}=0$ for $\mu <V/2$ and changes dramatically
around $\mu =V/2$. The dramatic change of $\alpha _{xy}$ disappears at
medium temperatures, but the sign change still remains (see Fig. \ref%
{bilayer}c). At different chemical potentials, the temperature dependence of
$\alpha _{xy}$ shows quite different behavior (see Fig. \ref{bilayer}d). For
uniform chemical potential at the valleys, the total anomalous Nernst signal
is zero, even though, like in the case of single layer graphene, they can be
substantial locally in the momentum space.

\subsection{Conclusion}

In summary, we discuss the effects of the topological Berry phases on the
thermoelectric response coefficients of the Dirac-like quasiparticles in
gapped single and bilayer graphene. The gap in the single layer graphene can
be due to SI symmetry breaking, as claimed, for instance, in epitaxial
graphene in Ref.~\cite{Zhou}. It may also be due to confined geometries,
such as in graphene nanoribbons \cite{Kim1,Sarma}. For bilayer graphene, the
gap may be due to applied voltage bias \cite{Ohta}. We show that in such
samples the anomalous contribution to the Nernst effect is non-zero for a
non-zero valley polarization. We also show that the nonlinear quasiparticle
dispersion spectrum does not significantly change the results. A non-zero
valley polarization may be artificially created \cite{Rycerz}, or it can be
induced by a perpendicular magnetic field due to the substantial valley
magnetic moment \cite{Xiao1} in a generic Nernst set up. The regular Nernst
effect in gapped graphene may, therefore, have a substantial anomalous
contribution because of the valley polarization. Even in the absence of any
valley polarization, the spontaneous Nernst effect from an individual valley
is locally non-zero, opening up the possibility to create a measurable
valley polarization on the transverse edges of a sample which can be
technologically very important.

\textbf{Acknowledgement}: Zhang is supported by WSU startup funds. Tewari is
supported by DOE/EPSCoR Grant \# DE-FG02-04ER-46139 and Clemson University
start up funds.


\end{document}